\documentclass[twocolumn,amsmath,amssymb,floatfix,prl,epsf,showpacs]{revtex4}
\usepackage{amsmath,amssymb,natbib,bm,graphicx,url,epsfig}
\usepackage[ansinew]{inputenc}

\newcommand{\be}{\begin{equation}}
\newcommand{\ee}{\end{equation}}
\newcommand{\bes}{\begin{equation}\begin{split}}
\newcommand{\ees}{\end{split}\end{equation}}

\begin{document}

\title{Smearing of the 2D Kohn anomaly in a nonquantizing magnetic field:
Implications for the interaction effects}

\author{T.~A.~Sedrakyan, E.~G.~Mishchenko, and M.~E.~Raikh}

\affiliation{Department of Physics, University of Utah, Salt Lake
City, UT 84112}

\date{\today}

\begin{abstract}
Thermodynamic and transport characteristics of a clean
two-dimensional interacting electron gas are shown to be sensitive
to the weak perpendicular magnetic field even at temperatures much
higher than the cyclotron energy, when the quantum oscillations
are completely washed out. We demonstrate this sensitivity for two
interaction-related characteristics: electron lifetime and the
tunnel density of states. The origin of the sensitivity is traced
to the field-induced smearing of the Kohn anomaly; this smearing
is the result of  curving of the semiclassical electron
trajectories in magnetic field.

\end{abstract}


\maketitle

{\em Introduction}. Consider a high-mobility 2D electron gas in a
nonquantizing perpendicular magnetic field, $B$, so that the
condition $k_{\mbox{\tiny F}}l\gg 1$, where $k_{\mbox{\tiny F}}$
is the Fermi momentum and $l=(\hbar c/eB)^{1/2}$ is the magnetic
length, is  met. It is a common knowledge that the thermodynamic
and transport characteristics  of the electron gas are strongly
(in oscillatory manner) sensitive to $B$ at temperatures
$T<\hbar\omega_c$, where $\hbar\omega_c=\hbar^2/ml^2$ is the
cyclotron quantum, and $m$ is the effective mass. It is also
commonly accepted that that the sensitivity to $B$ vanishes
rapidly as $T$ exceeds $\hbar\omega_c$. A natural question to ask
is whether electron-electron interactions in 2D can change the
situation. Previous studies (most recent \cite{martin03,adamov05})
suggest that the answer to this question is negative. The
magnitude of magneto-oscillations still falls off with $T$ as
$\exp(-2\pi^2T/\omega_c)$. The interaction-induced renormalization
of $\omega_c$ is singular in temperature, however this singularity
vanishes in the limit of a weak disorder.

In the present paper we demonstrate that, due to electron-electron
interactions, thermodynamic and transport characteristics of the
{\em clean} 2D gas remain sensitive to $B$ at  temperatures  much
{\em higher} than $\hbar\omega_c$. More specifically, we show that
the electron  self-energy, $\Sigma (\omega,B)$, contains a
correction which scales with $B$ as $(\omega/\epsilon_0)$, where
$\epsilon_0$ is expressed through the Fermi energy,
$E_{\mbox{\tiny F}}$, as
%$E_{\mbox{\tiny F}}=\hbar^2k_{\mbox{\tiny F}}^2/2m$ as
%defined as
\begin{equation}
\label{scale} \epsilon_0=\frac{2E_{\mbox{\tiny
F}}}{(k_{\mbox{\tiny F}}l)^{4/3}}=\hbar\omega_c(k_{\mbox{\tiny
F}}l)^{2/3},
\end{equation}
%and $E_{\mbox{\tiny F}}$ is the Fermi energy.
Eq. (\ref{scale}) sets the temperature scale $T\sim \epsilon_0\gg
\hbar\omega_c$.
%which is much bigger than $\omega_c$.
Then the corresponding spatial scale, $p_0^{-1}=(k_{\mbox{\tiny
F}}l)^{4/3}/k_{\mbox{\tiny F}}$,
%$p_0^{-1}=\hbar v_{\mbox{\tiny F}}/\epsilon_0$, where $v_{\mbox{\tiny F}}$
%is the Fermi velocity,
is much smaller than the Larmour radius, $R_{\mbox{\tiny
L}}=k_{\mbox{\tiny F}}l^2$.
%Then the electron motion in external potential is
%described by a characteristic momentum $R_{\mbox{\tiny
%L}}^{-1}=k_{\mbox{\tiny F}}/(k_{\mbox{\tiny F}}l)^2$, where
%$R_{\mbox{\tiny L}}$ is the Larmour radius. Momentum
%$R_{\mbox{\tiny L}}^{-1}$
%It is this momentum that
%appears in the description of interaction corrections to the
%tunnel density of states \cite{rudin97} and interaction-induced
%magnetotransport \cite{gornyi03}.
%and interaction corrections to the tunnel density of states.
%This is related to the physical picture of transport, diffusion of
%the Larmour circle, which underlies the interaction effects.
%In the present paper we demonstrate that there exists another
%characteristic scale of the momenta
%\begin{equation}
%\label{p0} p_0=\frac{k_{\mbox{\tiny F}}}{\left(k_{\mbox{\tiny
%F}}l\right)^{4/3}},
%\end{equation}
%which is germane for the interaction effects.
The easiest way to see the relevance of $p_0$ is to consider the
Friedel oscillations, $V_{\mbox{\tiny H}}(r)$, of electrostatic
potential created by a scatterer with a short-range  potential,
$U_{imp}({\bf r})$, in the presence of electron-electron
interaction, $V({\bf r}-{\bf r}_1)$. Away from the scatterer,
$V_{\mbox{\tiny H}}(r)$ modifies from $V_{\mbox{\tiny
H}}(r)\propto \sin(2k_{\mbox{\tiny F}}r)/r^2$ in a zero field to
%In particular, $p_0$
%quantifies the magnetic-field-induced
%lifting
%lifting of the periodicity of the Friedel oscillations. More
%precisely, consider a scatterer with a short-range  potential,
%$U_{imp}({\bf r})$,  Then the asymptotic behavior
%of  electrostatic potential, $V_{H}(r)$, away from the scatterer
%modifies from $V_H(r)\propto \sin(2k_{\mbox{\tiny F}}r)/r^2$
% in a zero field to
\begin{equation}
\label{modified} V_{\mbox{\tiny H}}(r) =-\frac{\nu_0g
V(2k_{\mbox{\tiny F}})}{2\pi r^2}\; \sin\Biggl[2k_{\mbox{\tiny
F}}r-\frac{(p_0r)^3}{12}\Biggr]\;,
\end{equation}
at finite $B$. Here $\nu_0=m/\pi\hbar^2$ is the free electron
density of states, $V(2k_{\mbox{\tiny F}})$ is the Fourier
component of $V({\bf r})$, and the parameter $g$ is defined as
\cite{rudin'97} $g=\int U_{imp}({\bf r})\;d{\bf r}$.
Eq.~(\ref{modified}) is valid within the domain $k_{\mbox{\tiny
F}}^{-1} \lesssim  r \lesssim R_{\mbox{\tiny L}}$, so that
$(p_0r)^3/12$ in the argument of sine does not exceed the the main
term, $2k_{\mbox{\tiny F}}r$. As follows from (\ref{scale}), the
momentum $p_0$ is intermediate between $R_{\mbox{\tiny L}}^{-1}$
and $k_{\mbox{\tiny F}}$.
%Correspondingly, the energy scale
%$\epsilon_0=\hbar v_{\mbox{\tiny F}}p_0$, where $v_{\mbox{\tiny
%F}}$ is the Fermi velocity, is intermediate between the the Fermi
%energy, $E_{\mbox{\tiny F}}$, and the cyclotron energy
%$\hbar\omega_c=E_{\mbox{\tiny F}}/N_{\mbox{\tiny F}}$, where
%$N_{\mbox{\tiny F}}=(k_{\mbox{\tiny F}}l)^2/2$ is the highest
%occupied Landau level.

% It is
%easy to see that $r \lesssim R_{\mbox{\tiny L}}$ is also the
%condition that the field induced term  $(p_0r)^3/12$ in the
%argument of sine in Eq.~(\ref{modified}) does not exceed the main
%term, $2k_{\mbox{\tiny F}}r$. This is the reflection of the fact
%that the value of momentum $p_0$ is intermediate between
%$R_{\mbox{\tiny L}}^{-1}$ and $k_{\mbox{\tiny F}}$, i.e.,
%$R_{\mbox{\tiny L}}^{-1}\ll p_0 \ll k_{\mbox{\tiny F}}$. The
%energy scale, related to $p_0$,
%\begin{equation}
%\epsilon_0=\frac{\hbar^2 k_{\mbox{\tiny
%F}}p_0}{m}=\frac{2E_{\mbox{\tiny F}}} {\left(k_{\mbox{\tiny
%F}}l\right)^{4/3}},
%\end{equation}
%where $E_{\mbox{\tiny F}}$ is the Fermi energy,  exceeds the
%cyclotron quantum, $\hbar\omega_c=E_{\mbox{\tiny
%F}}/N_{\mbox{\tiny F}}$, where $N_{\mbox{\tiny F}}=(k_{\mbox{\tiny
%F}}l)^2/2$ is the highest occupied Landau level.

%The underlying reason for modification of the Friedel oscillations
%in perpendicular magnetic field is that

The argument of Eq.~(\ref{modified}) can be inferred from the
simple qualitative consideration.
%%%%%%%%%%%%%%%%%%%%%%%%%%%%%%%%%%%%%%%%%%%%%%%%%%%%%%%%%%%%%%%%%%
%The form of the Friedel oscillations
%in the magnetic field can be inferred from the following
%qualitative consideration. The key point is that
%the
Classical trajectory of an electron in a weak magnetic field is
{\em curved}
%due to the Larmour motion%
even at the spatial scales  much smaller than $R_{\mbox{\tiny
L}}$.
%Then,
Due to this curving, the electron propagator, $G({\bf r}_1,{\bf
r}_2)$, between the points ${\bf r}_1$ and ${\bf r}_2$
%exact Green function, $G({\bf r}_1,{\bf r}_2)$,
%in a weak magnetic field
contains,  in the semiclassical limit, a  phase $\hbar^{-1}S({\bf
r}_1,{\bf r}_2)=k_{\mbox{\tiny F}}{\cal L}$, where ${\cal L}$ is
the length of the arc of a circle with the radius
%equal to
$R_{\mbox{\tiny L}}$, that connects the points ${\bf r}_1$ and
${\bf r}_2$, see Fig.~1a. Since the Friedel oscillations are
related to the propagation from ${\bf r}_1$ to ${\bf r}_2$ {\em
and back},
%polarization operator contains the product of
%two such Green functions,
it is important that two arcs,
 corresponding to the opposite directions of propagation, define a {\em finite} area,
$\cal A$, so that the product $G({\bf r}_1,{\bf r}_2)G({\bf
r}_2,{\bf r}_1)$ should be multiplied by the Aharonov-Bohm phase
factor, $\exp\left[iB{\cal A}/\Phi_0\right]$. Then the phase,
$\varTheta$, of this product is equal to
\begin{equation}
\label{theta} \varTheta({\bf r}_1,{\bf r}_2)=\frac{2}{\hbar}S({\bf
r}_1, {\bf r}_2)-\frac{B{\cal A}({\bf r}_1,{\bf r}_2)}{\Phi_0}.
\end{equation}
%Fig. 4a. Then the Aharonov-Bohm flux, $B{\cal
%A}/\Phi_0$, through this area must be included into the phase,
%$\Theta$, corresponding to the polarization operator, i.e.,
%\begin{equation}
%\label{theta} \Theta({\bf r}_1,{\bf r}_2)=\frac{2}{\hbar}S({\bf
%r}_1, {\bf r}_2)-\frac{B{\cal A}({\bf r}_1,{\bf r}_2)}{\Phi_0}.
%\end{equation}
From simple geometrical relations $r=\vert {\bf r}_1-{\bf
r}_2\vert=2R_{\mbox{\tiny L}}\sin(\delta/2)$, ${\cal
L}=R_{\mbox{\tiny L}}\delta$ and ${\cal A}=2 R_{\mbox{\tiny
L}}^2(\delta-\sin\delta)$, we find for $r \ll R_{\mbox{\tiny L}}$
that
 $\varTheta=2k_{\mbox{\tiny
F}}r-(p_0r)^3/12$,  which coincides with
%the phase of
the argument in Eq.~(\ref{modified}).
%The scale $p_0$ can be inferred
%from a simple qualitative consideration. In the real space
%polarization is related to the product $G({\bf
%r,r^{\prime}})G({\bf r^{\prime},r})$ integrated over ${\bf
%r^{\prime}}$. For the spatial scale $p_0^{-1}=\rho_0$, the points
%${\bf r}$ and ${\bf r^{\prime}}$ are separated by $\sim \rho_0$.
%Then relevant Feynman trajectories are confined to the
%cigar-shaped area with axes $\rho_0$ and $(\rho_0/k_{\mbox{\tiny
%F}})^{1/2}$. Then the condition that the magnetic flux through the
%area $\rho_0(\rho_0/k_{\mbox{\tiny F}})^{1/2}$ is equal to the
%flux quantum yields that $(\rho_0)^{-1}$ is given by Eq.~%(\ref{p0}).
We emphasize, that the conventional way~\cite{gorkov59} of
incorporating magnetic field into the semiclassical Green's
function neglects the curvature of the electron trajectories.
%Therefore, this
%Such an incorporation
This incorporation would not capture the modification
(\ref{modified}) of the Friedel oscillations.

In the subsequent sections  we trace how the scale, $p_0$,
originates from the smearing of the Kohn anomaly in a magnetic
field, and explore the consequences of this smearing for two
interaction-related characteristics of the electron gas: electron
lifetime and the tunneling density of states in the ballistic
regime.

%%%%%%%%%%%%%%%%%%%%%%%%%%%%%%%%%%%%%%%%%%%%%%%%%%%%%%%%%%%%%%%%
\begin{figure}[t]
\centerline{\includegraphics[width=80mm,angle=0,clip]{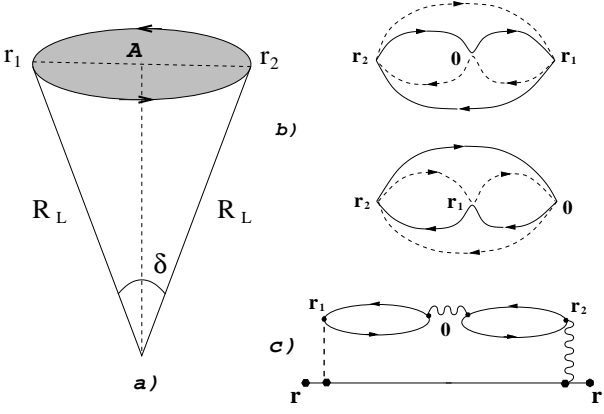}}
\caption{(a)  Origin of the ``magnetic'' term in the argument
Eq.~(\ref{modified}): two arcs, corresponding to the opposite
directions of propogation, define a finite area, ${\cal A}$; (b)
Three-scattering processes responsible for magnetic field
dependence of the relaxation rate are illustrated schematically;
(c) One of three RPA-type diagrams describing  the $B$-dependent
correction to the density of states in the ballistic regime:
scattering by the impurity occurs at point ${\bf r}_1$; two
electron-electron scattering acts occur at points $\{0,{\bf
r}_2\}$.}
\end{figure}
%%%%%%%%%%%%%%%%%%%%%%%%%%%%%%%%%%%%%%%%%%%%%%%%%%%%%%%%%%%%%%%%%
%In the remainder of the paper we derive
%Eqs.~(\ref{p0}-\ref{AiBi}), generalize them to finite temperatures
%and frequencies, and demonstrate the consequences of the
%field-induced smearing of $\Pi^{\prime}$ for two particular
%examples: electron energy relaxation in a clean 2D gas,  and the
%tunneling density of states in the ballistic regime.

%Within this accuracy,
%the scale, $\epsilon_0$,
%for the smearing of the Kohn anomaly
%would not emerge.
%%%%%%%%%%%%%%%%%%%%%%%%%%%%%%%%%%%%%%%%%%%%%%%%%%%%%%%%%%%%%%%%%

\noindent{\em Polarization operator.} Curving of the classical
electron trajectory in a weak magnetic field smears  the Kohn
anomaly in the polarization operator, $\Pi(q)$. On the
quantitative level, it is convenient to describe this smearing for
the derivative $\Pi^{\prime}(q) = d \Pi(q)/dq$. In a zero field
and at zero temperature, $\Pi^{\prime}(q)$ is equal to
$-(m/2\pi)\Theta(\delta q)/(\delta q k_{\mbox{\tiny F}})^{1/2}$,
where $\delta q =(q-2k_{\mbox{\tiny F}})$, and $\Theta(x)$ is the
step-function \cite{stern67}. As shown below, for a nonquantizing
field,
%$d \Pi/dq$
$\Pi^{\prime}(q)$ assumes the form
\begin{equation}
\label{AiBi}
%\frac{d \Pi}{dq}
\Pi^{\prime}(q) =-\frac{m}{(k_{\mbox{\tiny
F}}{p_0})^{1/2}}\;{\text{\large \it Ai}}\left(\frac{\delta
q}{p_0}\right) \;{\text{\large \it Bi}}\left(\frac{\delta
q}{p_0}\right),
\end{equation}
where $Ai(z)$ is the Airy function, and  $Bi(z)$ is another
solution of the Airy equation defined, {\em e.g.}, in Ref.
\onlinecite{book}. The rhs of Eq.~(\ref{AiBi}) is plotted in
Fig.~2. It is seen that the singularity at $q=2k_{\mbox{\tiny F}}$
is smeared by the magnetic field in a rather peculiar way: for
positive $\delta q \gg p_0$ the $(\delta q)^{-1/2}$-behavior is
restored. However, for large negative $\delta q/p_0$, the
derivative
%$d\Pi/dq$
$\Pi^{\prime}(q)$ approaches zero {\em with oscillations}, namely,
as $\cos\bigl[4(\vert \delta
q\vert/p_0)^{3/2}/3\bigr]/(\vert\delta q\vert)^{1/2}$.
%Such an
These oscillations
%oscillatory behavior
%is directly related to
%has
have the same  physical origin as the term $(p_0r)^3/12$ in the
phase of the Friedel oscillations (\ref{modified}).

\noindent{\em Derivation of Eq. (\ref{modified})}.
%To reveal the scale $(q-2k_{\mbox{\tiny
%F}})\sim p_0$ it is most convenient to
We start from the general expression \cite{aleiner95} for the
polarizability in a magnetic field
\begin{eqnarray}
\label{general} {\large{\Pi}}\;
(q)=-\frac{2m}{\pi}\sum_{n_1=0}^{\infty}\sum_{n_2=0}^{\infty}
\frac{(-1)^{(n_2-n_1)}\bigl(f_{n_1}-f_{n_2}\bigr)}{{n_2-n_1}}\nonumber\\
\times
%e^{(-q^2l^2/2)}
\exp(-q^2l^2/2)\;{\text{\large
L}}_{n_1}^{n_2-n_1}\left(\frac{q^2l^2}{2}\right) {\text{\large
L}}_{n_2}^{n_1-n_2}\left(\frac{q^2l^2}{2}\right),
\end{eqnarray}
where ${\text{\large L}}_{n_1}^{n_2-n_1}(x)$ and ${\text{\large
L}}_{n_2}^{n_1-n_2}(x)$ are the Laguerre polynomials, and
$f_n=\bigl\{\exp\bigl[(n-N_{\mbox{\tiny
F}})\hbar\omega_c/T\bigr]+1\bigr\}^{-1}$, with $N_{\mbox{\tiny
F}}=E_{\mbox{\tiny F}}/\hbar\omega_c$, is the Fermi distribution.
At small $q\ll k_{\mbox{\tiny F}}$ Eq.~(\ref{general}) yields
\cite{aleiner95} $\Pi\;\!(q)=-(m/\pi)\bigl[1-J_0^2(qR_{\mbox{\tiny
L}})\bigr]$, i.e., the characteristic scale is $q\sim
R_{\mbox{\tiny L}}^{-1}$. For $(q-2k_{\mbox{\tiny F}})\ll
k_{\mbox{\tiny F}}$ it is convenient to perform the summation over
the Landau levels with the help of the following integral
representation of the Laguerre polynomial
\begin{eqnarray}
\label{representation} {\text{\large
L}}_m^n(x)=\frac{1}{2\pi}\int_0^{2\pi}\!\!\frac{d\theta}{\left(1-e^{i\theta}\right)^{n+1}}
\exp\left\{\frac{xe^{i\theta}}{e^{i\theta}-1}-im\theta\right\}
\end{eqnarray}
%The small factor $\exp(-q^2l^2/2)$ in Eq~(\ref{general}) is
%compensated by the exponent in Eq.~(\ref{representation}).
In the vicinity $q=2k_{\mbox{\tiny F}}$ Eq.~(\ref{representation})
contains a small factor $\exp(-q^2l^2/2)$. This factor is
compensated by the product of the Laguerre polynomials, since each
of them is $\propto\exp(x/2)$, which comes from the exponent in
Eq.~(\ref{representation}) taken at $\theta=\pi$. With
contribution from the vicinity of $\theta\sim\pi$ dominating the
integral (\ref{representation}), we can expand the integrand
around this point as $\exp\bigl[x/2+i\pi m
+i\phi(\psi)\bigr]/2^{n+1}$, where $\psi=(\theta - \pi)$, and the
phase, $\phi(\psi)$, is
%equal to
\begin{eqnarray}
\label{expansion} {\large\phi}(\psi)= \left({x\over
4}-m-\frac{n+1}{2} \right)\psi+\frac{x\psi ^3}{48}.
\end{eqnarray}
%The key step is the summation over Landau levels in
%Eq.~(\ref{representation}).
%%%%%%%%%%%%%%%%%%%%%%%%%%%%%%%%%%%%%%%%%%%%%%%%%%%%%%%%%%%%%%%
\begin{figure}[t]
\centerline{\includegraphics[width=90mm,angle=0,clip]{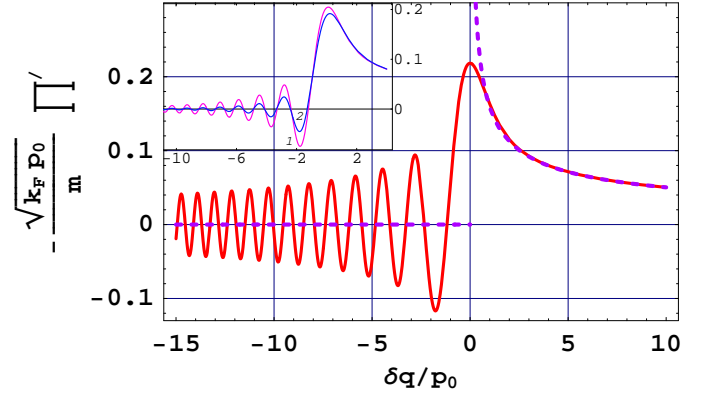}}
\caption{(Color online) Derivative, $\Pi^{\prime}$, of the
polarization operator is plotted from Eq.~(\ref{AiBi}) vs.
dimensionless deviation $\delta q/p_0=(q-2k_{\mbox{\tiny F}})/p_0$
from the Kohn anomaly. Dashed line shows $\Pi^{\prime}$ in  a zero
magnetic field. Inset: $\Pi^{\prime}$ is plotted from
Eq.~(\ref{finiteT}) at temperatures: $2\pi T=\epsilon_0$ (1),
 and $2\pi T=1.5\epsilon_0$ (2).} \label{fig:1}
\end{figure}
%%%%%%%%%%%%%%%%%%%%%%%%%%%%%%%%%%%%%%%%%%%%%%%%%%%%%%%%%%%%%%%%%
Now we make use of the fact that only relatively small number
$\sim (k_{\mbox{\tiny F}}l)^{2/3}\ll N_{\mbox{\tiny F}}$
%\frac{E_{\mbox{\tiny F}}}{\hbar
%\omega_c}=\frac{(k_{\mbox{\tiny F}}l)^2}{2}$
 of Landau levels around
$E_{\mbox{\tiny F}}$ contribute to the sum
Eq.~(\ref{representation}). This suggests that we can present
$n_1$ and $n_2$ as $n_1=N_{\mbox{\tiny F}}+m_1$ and
$n_2=N_{\mbox{\tiny F}}-m_2$, respectively,  and extend the sum
over $m_1$, $m_2$ from $-\infty$ to $+\infty$. After that, the
summation over Landau levels can be easily carried out with the
help of the
%following
identity
\begin{eqnarray}
\label{identity2} \sum\limits _{m_1,m_2=-\infty}^{\infty}
\frac{f_{N_{\mbox{\tiny F}}-m_1}-f_{N_{\mbox{\tiny
F}}+m_2}}{m_1+m_2}
%\times\qquad\qquad\nonumber\\
\cos\bigl[(m_1-m_2)\alpha+\beta\bigr]\nonumber\\
=\frac{2 \pi ^2T\cos\beta}{\hbar \omega _c\sinh \bigl(2\pi\vert
\alpha\vert T/ \hbar\omega _c)}.\qquad\qquad\qquad
\end{eqnarray}
%\begin{eqnarray}
%\label{identity1}
%\sum_{m_1,m_2>0}\frac{\cos(m_1\alpha)\cos(m_2\alpha)}{m_1+m_2}\qquad\qquad\qquad
%\nonumber\\
%=\sum_{m_1,m_2>0}
% \frac{\sin(m_1\alpha)\sin(m_2\alpha)}{m_1+m_2}=
%\frac{\pi}{4\vert \alpha \vert}.
%\frac{\varphi-\psi}{4}\tan\left(\frac{\varphi-\psi}{2}\right)\quad\\
%+\frac{\varphi+\psi}{4}\tan\left(\frac{\varphi+\psi}{2}\right)+
%\frac{1}{2}\ln\Bigl[(1-\cos\varphi)(1-\cos\psi)\Bigr],\nonumber
%\end{eqnarray}
Upon substituting Eq.~(\ref{representation}) into
Eq.~(\ref{general}) and using Eq.~(\ref{identity2}), one of the
angular integrations can be performed:
%explicitly, yielding
\begin{eqnarray}
\label{finiteT} \Pi^{\prime}(q,T)=&-&\frac{mT}{2^{1/6}(\pi
k_{\mbox{\tiny F}}{p_0})^{1/2}\epsilon_0}
\int_0^{\infty}\frac{dx\; x^{1/2}}{\sinh(2\pi xT/\epsilon_0)}\nonumber\\
&\times& \sin\left(2^{2/3}\frac{\delta
q}{p_0}\;x+\frac{1}{3}x^3+\frac{\pi}{4}\right)
%+ \cos\left(2^{2/3}\frac{\delta
%q}{p_0}\;x+\frac{1}{3}x^3\right)\Biggr].
\end{eqnarray}
It can be shown that in the limit $T\rightarrow 0$ the integral
(\ref{finiteT}) reduces to the product $Ai\;(\delta q/p_0)\cdot
Bi\;(\delta q/p_0)$, which falls off as $1/(\delta \vert q
\vert)^{1/2}$ and oscillates, see Fig.~2. As the difference
$2k_{\mbox{\tiny F}}-q$ increases and becomes comparable to
$k_{\mbox{\tiny F}}$,
%up to $\delta q
%\sim k_{\mbox{\tiny F}}$,
these oscillations cross over to the ``classical'' oscillations
$\Pi^{\prime}(q) \propto J_0(qR_{\mbox{\tiny
L}})J_1(qR_{\mbox{\tiny L}})\propto \cos(2qR_{\mbox{\tiny L}})$.

The behavior of $\Pi^{\prime}(q,T)$ at finite temperatures is
shown in Fig.~2. It is seen that  increasing $T$ suppresses the
oscillatory behavior of $\Pi^{\prime}(q,T)$ starting from $2\pi
T\approx 1.5\epsilon_0$. Now the large-distance behavior of the
potential, created by the short-range impurity, can be expressed
directly through $\Pi ^\prime (2k_{\mbox{\tiny F}}+ Q)$
% is related to $\Pi ^\prime (q)$ by the following
as follows

%%%%%%%%%%%%%%%%%%%%%%%%%%%%%%%%%%%%%%%%%%%%%%%%%%%%%%%%%%%%
\begin{eqnarray}
\label{calculation}
 V_{\mbox{\tiny H}}(r)
=\frac{V(2k_{\mbox{\tiny F}})g}{2(\pi k_{\mbox{\tiny F}}r)^{3/2}}
\int_{-\infty}^{\infty}\!\!dQ \sin\! \left[(2k_{\mbox{\tiny
F}}+Q)r-\!{\pi\over
 4}\right]\Pi ^\prime (Q,T).\nonumber
\end{eqnarray}
%Substituting (\ref{finiteT}) into (\ref{calculation}) and
%performing elementary integration, we obtain
%\begin{eqnarray}
%\label{T-mod} V_H(r) =-\frac{\nu_0g V(2k_{\mbox{\tiny
%F}})mT}{\hbar k_{\mbox{\tiny F}}r}
%\;\;\frac{\sin\Bigl[2k_{\mbox{\tiny
%F}}r-\frac{1}{12}(p_0r)^3\Bigr]}{\sinh\bigl(2\pi mrT/\hbar
%k_{\mbox{\tiny F}} \bigr)}\;.
%\end{eqnarray}
%%%%%%%%%%%%%%%%%%%%%%%%%%%%%%%%%%%%%%%%%%%%%%%%%%%%%%%%%%%%%%%
In the limit $T\rightarrow 0$ we immediately recover
Eq.~(\ref{modified}).

Note that, in 3D,
%the modification of
the Friedel oscillations in magnetic field
%was
were addressed in a number of papers spanning almost four decades
\cite{rensink68,glasser69,horing69,horing04}. However, the authors
of Refs.~\onlinecite{rensink68,glasser69,horing69,horing04} were
unable to carry out
%perform
the summation over Landau levels in the 3D version of
Eq.~(\ref{general}).
%the focus of the study was the anisotropy of the  Friedel
%oscillations, which is specific for three dimensions. Concerning
%the
% smearing of
%the logarithmic Kohn anomaly, it requires analytical summation
%over Landau levels in 3D version of Eq.~(\ref{general}); the
%authors of
%Refs.~\onlinecite{rensink68,glasser69,horing69,horing04} were
%unable to carry it out.

%With $\Pi ^\prime (q_1)$ given by Eq.~(\ref{general}) we
%immediately recover Eq.~(\ref{modified}) at $T=0$.
%At finite
%temperature the Friedel oscillations are damped. However, the
%substantial damping occurs only for $T\gg \epsilon _0$, as
%illustrated in Fig.~2.

%Following the above reasoning,
The above derivation of the polarization operator near
$q=2k_{\mbox{\tiny F}}$ can be easily extended to finite
frequencies. For simplicity we present only a zero-temperature
result
\begin{eqnarray}
\label{finite_o} \Pi ^\prime
(\omega,q)\!&=&\!\frac{m}{2(k_{\mbox{\tiny
F}}p_0)^{1/2}}\sum_{s=+,-}\Biggl[ -{\text{\large \it
Ai}}\left(\frac{\delta q_s}{p_0}\right) \;{\text{\large \it
Bi}}\left(\frac{\delta
q_s}{p_0}\right)\nonumber\\
\!&+&\!\frac{2\omega}{v_{\mbox{\tiny F}}p_0}{\text{\large \it
Ai}}\left(\frac{\delta q_s}{p_0}\right) \;{\text{\large \it
Ai}}\;^{\prime}\left(\frac{\delta q_s}{p_0}\right) \Biggr],
\end{eqnarray}
\vspace{4mm} where $\delta q_{\pm}$ are defined as $\delta
q_{\pm}=(\delta q \pm \omega/v_{\mbox{\tiny F}})$. It is
straightforward to check that in the limit $l\rightarrow \infty$
Eq.~(\ref{finite_o}) reproduces the zero-field result
\cite{stern67} $\Pi ^\prime (q,\omega) \propto [\left(\delta
q_{+}\right)^{-{1/ 2}}+ \left(\delta q_{-}\right)^{-{1/ 2}}]$.
%\newline

\noindent{\em Implications}. The two most prominent
characteristics of a clean electron gas, that are governed by the
Kohn anomaly, are the energy dependence of the electron lifetime,
 $\tau_{e}^{-1}$, and the nonanalytic
correction, $\delta C(T)={\tilde \gamma}T^2$, to the Fermi-liquid
expression, $C(T)=\gamma T$, for the specific heat. Both
quantities are being extensively studied
\cite{millis,chubukov03,chubukov05,saha,suhas,Jungwirth96,zheng96,vignale05}.
Below we demonstrate that, in a weak magnetic field,
$\tau_{e}^{-1}$ acquires a correction, which is singular in $B$.
The standard expression \cite{Jungwirth96,zheng96}
$\tau_{e}^{-1}(\omega)=\left(1+f^2_s\right)\left(\omega^2/4\pi
E_{\mbox{\tiny F}}\right)\ln\left(E_{\mbox{\tiny
F}}/\omega\right)$ is derived within the random-phase
approximation (RPA). Here $f_s=r_s/\left(r_s+\sqrt{2}\right)$, and
$r_s$ is the interaction parameter. The RPA definition of
$\tau_{e}^{-1}$ can be conveniently cast in the form
%\vspace*{-0.2truein}
\begin{eqnarray}
\label{moment_rpa}
\frac{1}{\tau_e(\omega)}&=&\frac{4}{\nu_0}\int_{0}^{\omega}\!\!{d\Omega}\!\!\int
\frac{d^2{q}\;d^2{p}}{(2\pi)^4}A(\Omega,{\bf p})A(\omega,\vert{\bf
q}+{\bf
p}\vert)\nonumber\\
&\times &{\text{Im}}\left[\frac{1}{\Pi\left(\omega-\Omega,{\bf
q}\right)-q\nu_0/(\sqrt{2}k_{\mbox{\tiny F}}r_s)}\right],\qquad
\end{eqnarray}
%\vspace{-2mm}
where $A(\omega,{\bf q})$ is the spectral function related to the
retarded, $G^R$, and advanced, $G^A$, Green functions in a
standard way: $A=\left(G^R-G^A\right)/2\pi i$. In a zero magnetic
field, the product of the spectral functions in
Eq.~(\ref{moment_rpa}) accounts for the momentum and energy
conservation. The $B$-dependence of $\tau_{e}^{-1}$ comes from the
momenta close to $2k_{\mbox{\tiny F}}$ in (\ref{moment_rpa}).
Hence, to extract this dependence,  one should expand the
polarization operator in the denominator of Eq.~(\ref{moment_rpa})
in the vicinity of $2k_{\mbox{\tiny F}}$, and employ
Eq.~(\ref{finite_o}). However, the first order expansion {\em does
not} result in the $B$-dependence of $\tau_{e}^{-1}$. This happens
due to the cancellation of the $B$-dependent contributions coming
from the polarization operator and from the product of the
spectral functions.
%which, similarly to
%$\Pi(\omega,q)$, reduces to the products of the Green functions.
%As a result of
Because of this cancellation, the $B$-dependence of
$\tau_{e}^{-1}$ emerges upon expansion of the denominator in
(\ref{moment_rpa}) up to the {\em second} order in
$\tilde{\Pi}(q,\omega)=\left\{\Pi(q,\omega)-\Pi(2k_{\mbox{\tiny
F}},\omega)\right\}$
%according to
which amounts to replacement
$[\Pi(q,\omega)-q\nu_0/\sqrt{2}k_{\mbox{\tiny
F}}r_s]^{-1}\rightarrow -{\tilde \Pi}^2(q,\omega)f_s^3/\nu_0^3$.
%where ${\tilde \Pi}(q,\omega)$ is the
%contribution to the polarization operator coming from the domain
%of momenta close $2k_{\mbox{\tiny F}}$.
It is much more convenient to present the result of this expansion
in the coordinate rather than in the momentum space, where
$B$-dependent contribution to the relaxation rate  assumes the
following form
\begin{widetext}
%\vspace*{-0.2truein} \noindent
%\hrulefill \hspace*{3.6truein}
\begin{eqnarray}
\label{polar} \bigl\{\tau_e^{(2)}(\omega,B)\bigr\}^{-1}
=-\frac{4f_s^3}{\nu_0^4}\int_0^{\omega}{d \Omega}\int d{\bf
r_1}d{\bf r_2}A(\Omega,\vert {\bf r}_1-{\bf
r}_2\vert)A(\omega,\vert {\bf r}_1-{\bf
r}_2\vert)\;{\text{Im}}\Bigl[\tilde\Pi\left(\omega-\Omega,{\bf
r_1}\right)\tilde\Pi\left(\omega-\Omega,{\bf r_2}\right)\Bigr],\qquad\\
{\text{where}}\quad A(\omega,{\bf r})=\frac{\nu_0}{(2\pi
k_{\mbox{\tiny F}}r)^{1/2}} \sin\Biggl[k_{\mbox{\tiny
F}}r+\frac{\pi}{4}-\frac{(p_0r)^3}{24}+ \frac{\omega
r}{v_{\mbox{\tiny F}}}\Biggr]\quad  {\text{and}}\quad
\tilde\Pi\left(\Omega,{\bf r}\right)=-\frac{\nu_0}{2\pi r^2}
\sin\Biggl[2k_{\mbox{\tiny
F}}r-\frac{(p_0r)^3}{12}\Biggr]\exp{\left\{i\frac{\Omega
r}{{v_{\mbox{\tiny F}}}}\right\}}\quad\nonumber
\end{eqnarray}
\end{widetext}
\vspace{-0.2truein} \noindent
%\hspace*{-3.6truein}\noindent \hrulefill
are the spectral function in the coordinate space, and the inverse
Fourier transform of Eq.~(\ref{finite_o}), respectively. The major
contribution to the integral (\ref{polar}) comes from the domain
in which all three points $\{0,r_1,r_2\}$ are close to the
straight line (see Fig.~1b), so that the rapid oscillations in the
phase factors cancel each other. At $B=0$, this restriction causes
a relative smallness,
%$\sim~\!\omega^{5/2}/E_{\mbox{\tiny F}}^{1/2}$
by a factor $\sim ~\!(\omega/E_{\mbox{\tiny F}})^{1/2}$, of the
higher-order correction \cite{suhas} to the lifetime. At finite
$B$, the integrand in Eq.~(\ref{polar}) contains the exponent
$\exp\left\{i[\varTheta(\vert {\bf r}_1-{\bf
r}_2\vert)-\varTheta({r}_1)-\varTheta({ r}_2)]\right\}=
\exp[ip_0^3r_1r_2(r_1+r_2)/4]$; it is this exponent that is
responsible for the $B$-dependence of $\tau_{e}^{-1}$.
Essentially, at $\omega \lesssim \epsilon_0$, the small factor,
$(\omega/E_{\mbox{\tiny F}})^{1/2}$, caused by the angular
restriction, is replaced by $(\epsilon_0/E_{\mbox{\tiny
F}})^{1/2}=(2\omega_c/E_{\mbox{\tiny F}})^{1/3}$. The final result
for the correction to the lifetime can be presented as
\begin{equation}
\label{final} \frac{\delta\tau_{e}(\omega,B)}{\tau_{e}(\omega,0)}
=\left(\frac{2f_s^3}{1+f_s^2}\right)\frac{F(\omega/\epsilon_0)}{\ln(E_{\mbox{\tiny
F}}/\omega)}\;\left(\frac{2\omega_c}{E_{\mbox{\tiny
F}}}\right)^{1/3}\!\!\!,
\end{equation}
%where the
where $F(x)$ is the dimensionless function with the magnitude and
scale $\sim 1$. Note, that the correction (\ref{final}) is
strongly (as $B^{1/3}$) singular in $B$. The origin of this
singularity is the cubic dependence of the phase $\Theta$,
Eq.~(\ref{theta}), on the distance.

%A question can be raised about the higher order terms in the
%expansion of the polarization operator in the vicinity of
%$q=2k_{\mbox{\tiny F}}$, which we neglected. It can be
%demonstrated that the characteristic energy scale emerging from
%these terms is also $\epsilon_0$. However, these terms contain
%additional angular integrations, each adding a small factor
%$(\omega/E_{\mbox{\tiny F}})^{1/2}$ due to the fact that {\em all}
%points must be close to the straight line \cite{suhas}. Thus, the
%contributions off the higher-order terms fall off as powers of
%$(\omega/E_{\mbox{\tiny F}})^{1/2}$  with rising order.

The scale, $\epsilon_0$, also manifests itself in the zero-bias
anomaly in the tunnel density of states, $\nu(\omega)$, which is
closely related to the lifetime
%relaxation rate
(see, {\em e.g.}, Ref. \onlinecite{abrahams}). In the ballistic
regime $\omega \gg \tau^{-1}$, where $\tau$ is the elastic
scattering time \cite{rudin'97}, only the single-impurity
scattering processes determine the $\omega$-dependence of $\nu$.
The diagram, responsible for the $B$-dependence of the {\em local}
density of states (at point ${\bf r}$) in the ballistic regime, is
shown in Fig.~1c. It describes one impurity scattering at point
${\bf r}_1$ and two electron-electron scatterings at points
$\{0,{\bf r}_2\}$, and thus is quite similar to the process in
Fig.~1b. The difference in the analytic expressions for
$(\tau_{e}^{(2)})^{-1}$ and for the {\em average} RPA
$\delta\nu(\omega,B)$ is an extra integration over ${\bf r}$,
i.e.,
%$\delta\nu (\omega)=-\frac{2}{\pi}\;{\text{Im}}\;\int d{\bf
%r}\; \delta G_{\omega}({\bf r},{\bf r})$, where
\begin{eqnarray}
\label{deltaG} %\delta G_{\omega}\;({\bf r},{\bf r})
\delta\nu (\omega,B)= {\text{Im}}\frac{6if_s^2}{\pi^2 \nu_0^3
\tau}\int d{\bf r}\;d{\bf r}_1d{\bf r}_2\;G_{\omega}({\bf r},{\bf
r}_1)\qquad\qquad
\\\quad\qquad \times G_{\omega}({\bf r}_1,{\bf
r}_2)\tilde\Pi(0,{\bf r}_1)\tilde\Pi(0,{\bf r}_2)G_{\omega}({\bf
r}_2,{\bf r}).\nonumber
\end{eqnarray}
The factor $3$ in (\ref{deltaG}) reflects the fact that the
impurity scattering can occur not only at point ${\bf r}_1$ (as in
Fig.~1c), but also at points $0$ and ${\bf r}_2$.
%Eq.~(\ref{dos2})
Analysis of Eq. (\ref{deltaG}) yields
%Eq.~(\ref{int})
%yielding
%, and
%in the limit $\omega \ll \epsilon_0$ yields a similar result
\begin{equation}
\label{last}
\frac{\delta\nu(\omega,B)-\delta\nu(\omega,0)}{\nu_0}=
\frac{f_s^2\;F_1(\omega/\epsilon_0)}{E_{\mbox{\tiny F}}\tau}
\left(\frac{2\omega_c}{E_{\mbox{\tiny F}}}\right)^{1/3},
%1-\frac{F_1(0)}{\ln^2(E_{\mbox{\tiny F}}/\omega)}
%\left(\frac{\epsilon_0}{E_{\mbox{\tiny F}}}\right)^{1/2},
\end{equation}
where the function $F_1(x)$ is another dimensionless function with
scale and magnitude $\sim 1$.
%in the limit $\omega\ll\epsilon_0$ is a
%constant given by
%$F_1(0)=\frac{{3^{5/2}}\;\Gamma\left(5/6\right)\Gamma\left(2/3\right)}
%{2^{16/3}\;\pi\;\Gamma\left(7/6\right)}=0.203.$
Compared to the $B=0$ exchange correction \cite{rudin'97},
$\delta\nu(\omega,0)=-\nu_0\ln^2(E_{\mbox{\tiny F}}/\omega)/8\pi
E_{\mbox{\tiny F}}\tau$, Eq.~(\ref{last}) contains a small but
singular in $B$ factor $(2\omega_c/E_{\mbox{\tiny F}})^{1/3}$,
which has the same origin
%(angular restriction)
as  (\ref{final}).

% is the exchange part of the
%ballistic zero-bias anomaly \cite{rudin'97} at $B=0$; the expression for the
%constant $F_1(0)$ is the following $F_1(0)=
%\frac{{3^{3/2}\pi}\;\Gamma\left(5/6\right)\Gamma\left(2/3\right)}
%{2^{4/3}\;\Gamma\left(7/6\right)}=10.67$.
%In the limit $\omega \ll \epsilon_0$ the function $F_1$ yields the following constant
%$F_1(0)=\frac{{3^{3/2}}\;\Gamma\left(5/6\right)\Gamma\left(2/3\right)}
%{2^{13/3}\;\Gamma\left(7/6\right)}=0.425.$
\noindent{\em Conclusion}. A slight curving of the classical
trajectories in a weak magnetic field (as opposed to the drift of
the Larmour circle \cite{rudin97,gornyi03}), which had been
routinely disregarded since Ref.~\cite{gorkov59}, gives rise to
the singular corrections (\ref{final}), (\ref{last}) to the
lifetime and tunnel density of states, respectively. Although
these corrections are parametrically small in semiclassical
parameter $(k_{\mbox{\tiny F}}R_{\mbox{\tiny L}})^{-1}\ll 1$,
numerically, they turn out to be sizeable. For example, for
$B=0.2\;{\text T}$ and $n=2\cdot 10^{11}\;{\text cm}^{-2}$, the
parameter $(\epsilon_0/E_{\mbox{\tiny F}})^{1/2}$ is $\approx
0.5$. The scale, $\epsilon_0$, translates into a characteristic
temperature $T\sim\epsilon_0$, at which the lifetime is sensitive
to $B$. Since $\epsilon_0\gg\omega_c$, the effects of discreteness
of Landau levels $\propto \exp\left\{-2\pi^2 T/\omega_c\right\}$
are negligible at this $T$. Note, that even at $T\sim\epsilon_0$,
the inelastic  length, $v_{\mbox{\tiny F}}\tau_{e}(T)$, is $\sim
v_{\mbox{\tiny F}}E_{\mbox{\tiny F}}/
\epsilon_0^2\ln(E_{\mbox{\tiny F}}/\epsilon_0)$, i.e., it is
bigger than our characteristic spatial scale, $p_0^{-1}$, in
parameter $E_{\mbox{\tiny F}}/\epsilon_0\ln(E_{\mbox{\tiny
F}}/\epsilon_0)$.

Concerning experimental observability of the our predictions, we
note that $\tau_{e}$ at $B=0$ was extracted with  high accuracy
from the tunneling experiment \cite{murphy}, and was shown to be
consistent with
%correctly performed
RPA calculations \cite{Jungwirth96,zheng96}, (see also
\cite{vignale05}). We predict that sensitivity of the width of the
Lorentzian in the tunneling $dI/dV$ vs. $V$ characteristics to the
magnetic field persists down to low $B$.
%Up to now, the measurements of
%magnetotunneling between two clean
%electron layers were reported
%only for quantizing $B$ \cite{spielman}.

%$I$-$V$ characterisitcs to magnetic field
%persists down to low $B$.
%$\tau_{\epsilon}^{-1}\propto \Sigma^{\prime\prime}(\omega)$,
%the relaxation slows down in magnetic field, essentially by the factor
%$\ln\left(E_{\mbox{\tiny F}}/\omega\right)^2/\ln\left(E_{\mbox{\tiny F}}^2/\omega\epsilon_0\right)$
%for $\omega < \epsilon_0$. Note that, in the a zero magnetic field,  the rate, $\tau_{\epsilon}^{-1}$,

%In our consideration
%We had ignored the electron spin,
%on the basis that
%since the Zeeman splitting is $\sim \omega_c \ll \epsilon_0$ in 2D
%electron gas. Conversely, the spin response dominates for the gas
%of heavy carriers (such as $^3$He atoms)  \cite{chubukov05,saha}.

%However, for heavy carriers  (such as $^3$He atoms), $B$-dependent
%modification of nonanalytic terms in thermodynamic properties is
%governed by spin \cite{chubukov05,saha}.

Within the overall picture of 2D electron gas with disorder, the
energies $\omega < 1/\tau$ correspond to the diffusive regime,
while  energies $\omega > 1/\tau$ - to the ballistic regime. Our
main results Eqs. (\ref{final}), (\ref{last}) apply, when the new
energy scale, $\epsilon_0$, belongs to the ballistic domain, i.e.,
$\epsilon_0 \gg 1/\tau$. This quantifies our assumption that the
electron gas is clean. Under the same conditions, it can be
``dirty'', in the sense, that the mean free path can be much
smaller than the Larmour radius, i.e., $\omega_c\tau \ll 1$.
Numerical estimate shows that, {\em e.g.}, for $B= 0.2\;{\text T}$
the condition $\epsilon_0 \gg 1/\tau$ is met even for
%relatively
low mobilities $\sim 10^4\; {\text cm}^2/{\text{V s}}$.

Discussions with  A.~\!V.~\!Andreev,  A.~\!V.~\!Chubukov,
L.~\!I.~\!Glazman, B.~\!I.~\!Halperin,  D.~\!L.~\!Maslov, and
K.~\!A.~\!Matveev are gratefully acknowledged. E.G.M. acknowledges
the support of DOE under grant No. DE-FG02-06ER46313.

\end{document}